\documentclass[12pt]{amsart}

\usepackage{graphicx, verbatim,amsmath,amssymb, float}
\usepackage{hyperref}
\hypersetup{
    colorlinks=true,
    linkcolor=blue,
    filecolor=magenta,      
    urlcolor=cyan,
}

\def\vs#1 {\vskip#1truein}
\def\hs#1 {\hskip#1truein}

\newcommand{\real}{{\rm I\kern-.2em R}}

\newcommand{\be}{\begin{equation}}
\newcommand{\ee}{\end{equation}}

\newtheorem*{thm*}{Theorem}

\everymath{\displaystyle}

\theoremstyle{definition}

\begin{document}
\vs-.7
\hbox{}
\title[Random Close Packing and Hard Spheres]{Equivalence Between 
Random Close Packing in Granular Matter and
Freezing in the Hard sphere Model}

\date{\today}

\author{Charles Radin}
\address{Charles Radin\\Department of Mathematics\\The University of
Texas\\ Austin, TX 78712} \email{radin@math.utexas.edu}
\vs-.8
\hbox{}
\maketitle
\begin{abstract} 
The notion of random close packings of a
bulk static collection of ball bearings or sand grains was introduced
in the 1960's by G.D. Scott and J.D. Bernal. There have been numerous
attempts to understand the packings. We give a short argument,
based on recent experiments
and simulations, which explains the packings in purely geometric terms. 
\end{abstract}

In 1960 G.D. Scott and J.D. Bernal began a series of experiments, with
others, on large collections of static ball bearings and other
`granular matter'~\cite{Scott,BernalScott} in part studying how the
material compacts when shaken repeatedly. They discovered a barrier:
the system would not compact beyond a rather sharp volume fraction
they called `random close packing' (RCP), which they determined to be
about $0.64$.

 Many attempts have been made since then to illuminate
the phenomenon of shaken granular 
matter~\cite{ScottKilgour,Finney,Berryman,Aste,Baranau}.
The experiments started soon after the first
simulation to successfully exhibit the first order phase transition of
the hard sphere model of equilibrium statistical mechanics~\cite{Alder},
to which they refer and presumably used for intuition. In this paper
we extend some of Bernal's ideas about liquids~\cite{Bernal}
and use recent
experiments and simulations~\cite{Rietz,Jin} to show a close
connection between random close packing at volume fraction $0.64$ and
the freezing of a hard sphere fluid at density $0.49$~\cite{Loewen}, a
correspondence which we show is useful in both directions. We start
with a quick summary of the hard sphere model, and then  derive
a connection with random close packing.

The hard sphere model of particles in thermal equilibrium uses point
particles of mass $m$ constrained to be at least some fixed distance
$\sigma$ apart, with no other interaction. Given $\displaystyle x\in
\real^3$ and $s>1$ we introduce the constraint function $G_{x,s}$, of
variables $\displaystyle a,b,\cdots\in \real^3$, to have value 0 if
any of the variables $a,b,\cdots$ is further than $s$ from $x$, or if
any distinct pair of $x,a,b,\dots$ is closer than $\sigma$; otherwise
$G_{x,s}=1$. Fix (temperature) $T>0$, (pressure) $P>0$, and the number
$N+1$ of particles, and define the (temperature/pressure) relative
probability density that $N$ of the particles have position/momentum
coordinates $\displaystyle (x_1,p_1),(x_2,p_2),\cdots,(x_N,p_N)$ given
that the other has coordinates $(x,p)$, to be

\be
\exp\Big[-{1\over 2mkT}\Big({p\cdot p}+\sum_{j=1}^N {p_j\cdot
p_j}\Big)\Big]\int_0^\infty G_{x,s}(x_1,\cdots,x_N)\exp\Big[-{P\sigma^2\over
kT}\Big({4\pi s^3\over 3}\Big)\Big]ds.
\ee

This is the hard sphere model. Notice that the density
factors into a part dependent only on the momentum coordinates,
controlled by $T$, and a part dependent only on the position
coordinates, controlled by $P/kT$.

Simulation shows~\cite{Alder,Loewen} there is a unique infinite particle state for any
pair $(P,T)$ of pressure and temperature except for a special value
$\displaystyle {P\sigma^3/kT}=C^\ast$, $\displaystyle C^\ast$ a
dimensionless constant, where there is coexistence of the high density
phase which exists for $\displaystyle {P\sigma^3/kT}\ge C^\ast $ with
density $\ge 0.54$, and the low density phase which exists for
$\displaystyle {P\sigma^3/kT}\le C^\ast $ with density $\le 0.49.$
\vs-.2
\begin{figure}[H]
\center{\includegraphics[width=3in]{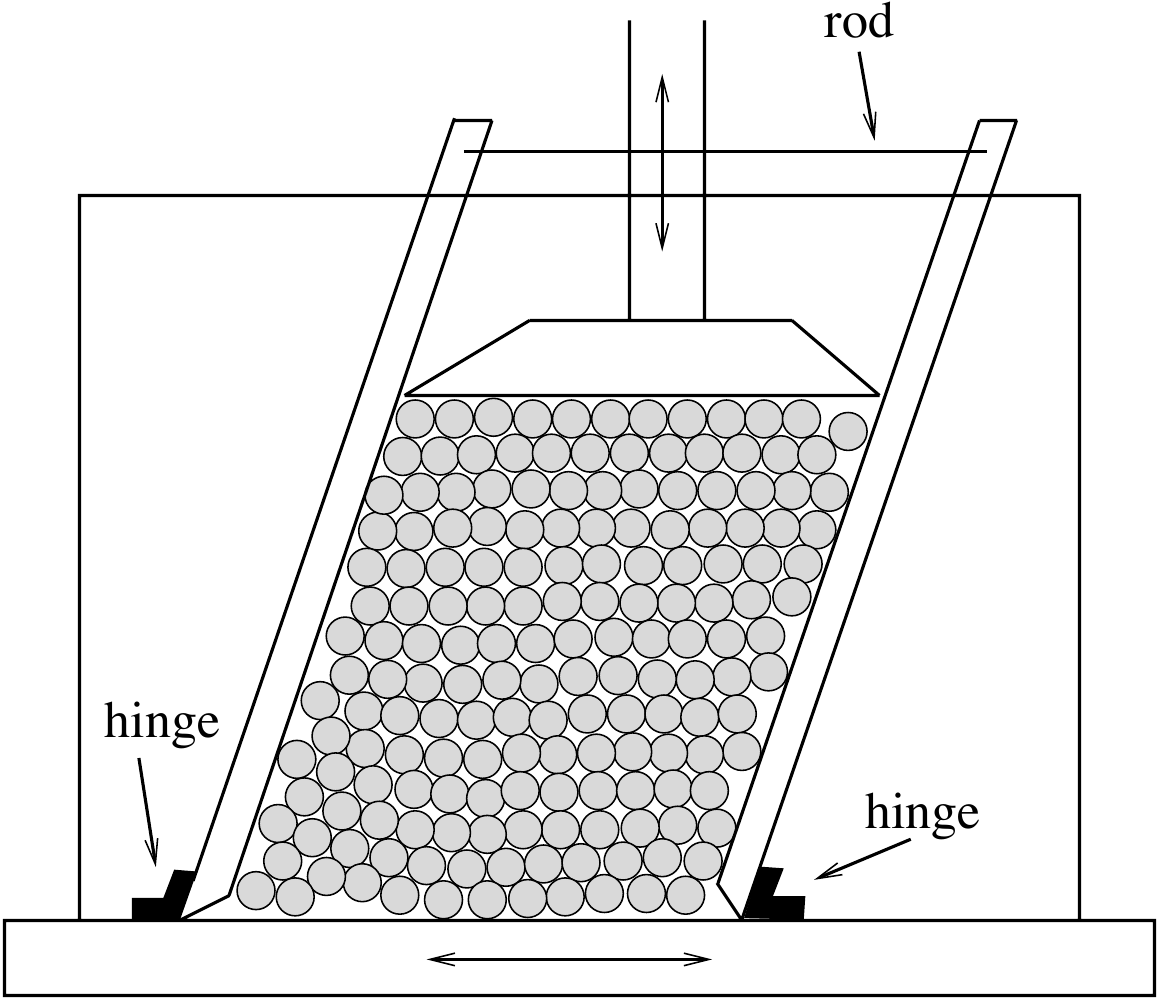}}
\vs.2
\caption{Sketch of a shear cell using two hinged vertical walls attached to a moving
bottom, and pressure applied from above.}
\label{shear-cell}
\end{figure}

The experiments in 1960 came in the midst of a long program of
Bernal's investigating the nature of liquids, in particular their
particle configurations; see~\cite{Finney-review} for a review. From
his 1959 paper~\cite{Bernal}, in which he concentrated on liquid
argon, we take the assumption that one can picture each argon atom in
the liquid as a hard ball of some diameter $\sigma$ vibrating at high
frequency, with a relatively slow drift, and that it is useful to
shift focus from the instantaneous ball positions to that of the
slowly moving vibrating balls, which we will call `clouds'. We add the
assumption that the clouds are roughly spherical and have a volume
which depends only on $(P,T)$, in particular that it is roughly the
same size in the coexisting liquid and solid states.

Thus one understands the microscopic origin of macroscopic properties
 of fluid argon to derive from essentially static clouds made of
 rapidly vibrating hard balls.  In particular these give rise to
 internal pressure of the liquid, supporting for instance the bulk
 modulus, by vibrating against neighboring clouds. Such clouds should
 be large enough to touch other clouds, to provide the pressure.  It
 is not a big leap for us to introduce clouds in the hard sphere
 model to model {\it static} sand grains, supporting the bulk modulus
 by contact forces rather than the vibration in the clouds.

In the crystalline state of the hard sphere model the size of the
cloud must be such as to touch the 12 neighbors,
giving a volume fraction about 0.74. Since the ratio of the volume
fractions in the coexisting states for the underlying vibrating
balls should be the same as the corresponding ratio for the clouds:
\be
{0.49 \over 0.54}\cong{\hbox{RCP}\over 0.74},
\ee
which yields a volume fraction for RCP about $0.67$, which is
within $4\%$ of the current best estimate of $0.646$~\cite{Rietz,Jin}.

In summary, the clouds of vibrating balls in thermal equilibrium
behave like well-mixed static sand grains, with the hard sphere
freezing point appearing as granular RCP at about 0.646. Indeed, one
might well use the hard sphere model to predict the existence of RCP
in sand, which one might say is what Bernal and Scott were exploring.

We now shift direction to examine the granular experiments and see
what they might say about the hard sphere model.

In 1964 Scott found a way to drive ball packings well above the
barrier of volume fraction $0.64$, and this was very
revealing~\cite{ScottCharlesworth,Nicolas,Rietz,Jin}. He found that if he cyclically
sheared a low volume fraction collection in a box with moving side
walls (a `shear cell'), as illustrated in Figure~\ref{shear-cell}, the
system would first compact to volume fraction 0.64, and then slowly
rearrange close to a densest packing at volume fraction 0.74.

\begin{figure}[H]
\center{\includegraphics[width=3in]{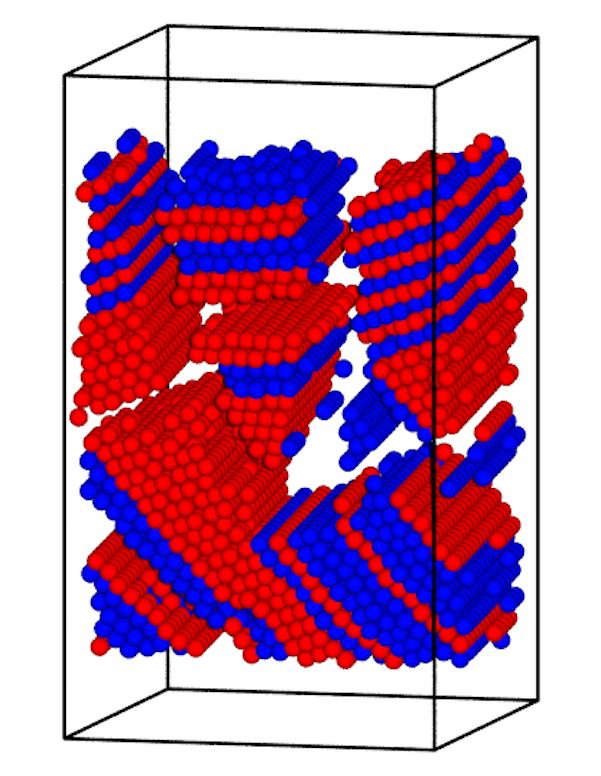}}
\caption{Hexagonal layers in growing granular crystallites, made from
unpublished simulation data by Jin for~\cite{Jin}. HCP layers are blue
and FCC layers are red.}
\label{layers}
\end{figure}

 There were difficulties understanding how the rearrangement took
place in time, since this would require looking inside the packing,
but this was recently done using laser sheets~\cite{Rietz,Jin} and
what was discovered is that in order for the packing to rearrange at
volume fraction $0.64$ it had to create small clusters of balls
(crystalline nuclei), randomly in the middle of the packing, which
then grew. This is, of course, how a liquid freezes (homogeneously),
so if one considers the nucleating granular system at volume fraction
$0.64$ as akin to a supercooled liquid about to freeze, its
correspondence with the hard sphere system at density $0.49$ suggests
that the phase in the hard sphere model at density above $0.54$ is not
just high density but `crystalline'.
Furthermore, examination of the grain configurations
from~\cite{Rietz,Jin}  shows that the high volume
fraction granular configurations consist of random layers of hexagonal
slabs; they are not regularly layered as would be FCC or
HCP. See Figure~\ref{layers}.
 Such a random layering might still have {\it orientational} long
range order from the direction normal to the layers, but not full
positional long range order as in a true crystal. The correspondence
with the hard sphere model suggests the same is true for the high
density phase in that model. (For a related question about hard
colloids see \cite{Pronk}.)

In summary, 
the evidence from the hard sphere model of a singularity in ball
packings as a function of volume fraction is remarkable.  The recent
evidence from~\cite{Rietz,Jin} that granular matter crystallizes
homogeneously is then suggestive that there is a simple connection to
freezing in the hard sphere model, so that there really exists only
one such ball packing singularity.


\begin{thebibliography}{1234}

\bibitem{Scott} G.D. Scott, Packing of Spheres, Nature (London) 188 (1960) 908-909.

\bibitem{BernalScott} J.D. Bernal and J. Mason, Co-ordination of
Randomly Packed Spheres, Nature (London) 188 (1960) 910-911.

\bibitem{ScottKilgour} G.D. Scott and D.M. Kilgour, The density of
random close packing of spheres. J. Phys. D 2, (1969) 863–866.

\bibitem{Finney} J.L. Finney,  Random packings and the structure of simple
liquids. I. The geometry of random close packing. Proc. R. Soc. A 319,
(1970) 479–493.

\bibitem{Berryman} J.G. Berryman, Random close packing of hard spheres
and disks, Phys. Rev. A 27 (1983) 1053–1061.

\bibitem{Aste} T. Aste, M. Saadatfar and T.J. Senden, Geometrical
structure of disordered sphere packings. Phys. Rev. E 7 (2005) 061302.

\bibitem{Baranau} V. Baranau, and U. Tallarek, Random-close packing
limits for monodisperse and polydisperse hard spheres, Soft Matter 10
(2014) 3826–3841.




\bibitem{Alder} B. Alder and T. Wainright, Phase transition for a hard
sphere system, J. Chem. Phys. 27, (1957) 1208-1209.

\bibitem{Bernal} J.D. Bernal, A geometrical approach to
  the structure of liquids, Nature (London) 183 (1959) 141-147.





\bibitem{Rietz} F. Rietz, C. Radin, H. Swinney and M. Schroeter,
Nucleation in sheared granular matter, Phys. Rev. Lett. 120 (2018)
055701.

\bibitem{Jin} W. Jin, C. O'Hern, C. Radin, M. Shattuck and H.L. Swinney, 
Homogeneous crystallization in cyclically sheared frictionless grains,
Phys. Rev. Lett. 125 (2020) 258003.

\bibitem{Loewen} H. Loewen, Fun with hard spheres. In: Mecke, K.R.,
Stoyen, D. (eds.)  Statistical Physics and Spatial Statistics: The Art
of Analyzing and Modeling Spatial Structures and Pattern
Formation. Lecture Notes in Physics, vol. 554, pp. 295–331. Springer,
Berlin (2000).

\bibitem{Finney-review} J.L. Finney, Bernal's road to random packing
  and the structure of liquids, Philos. Mag. 13 (2013) 3940-3969

\bibitem{ScottCharlesworth} G. D. Scott, A. M. Charlesworth and
M. K. Mak, On the Random Packing of Spheres, J. Chem. Phys. 40 (1964)
611-612.

\bibitem{Nicolas}
M.\ Nicolas, P.\ Duru and O.\ Pouliquen,
Compaction of a granular material under cyclic shear,
Eur. Phys. J. E {3} (2000) 309-314.

\bibitem{Pronk} S. Pronk and D. Frenkel, Can stacking faults in
hard-sphere crystals anneal out spontaneously?, J. Chem. Phys. 110
(1999) 4589-4592.


\end{thebibliography}
\end{document}